# A QoI Based Energy Efficient Clustering For Dense Wireless Sensor Network

MANJU PRASAD[#1], ANDHE DHARANI[#2]

[#]Department of Master of Computer Application, R.V College of Engineering,
Bangalore, India
[1]`manjuprasad32@gmail.com`, [2]`andhedharani@rvce.in`,

*ABSTRACT*

In a wireless sensor network Quality of Information (QoI), Energy Efficiency, Redundant data avoidance, congestion control are the important metrics that affect the performance of wireless sensor network. As many approaches were proposed to increase the performance of a wireless sensor network among them clustering is one of the efficient approaches in sensor network. Many clustering algorithms concentrate mainly on power Optimization like FSCH, LEACH, and EELBCRP. There is necessity of the above metrics in wireless sensor network where nodes are densely deployed in a given network area. As the nodes are deployed densely there is maximum possibility of nodes appear in the sensing region of other nodes. So there exists an option that nodes have to send the information that is already reached the base station by its own cluster members or by members of other clusters. This mechanism will affect the QoI, Energy factor and congestion control of the wireless sensor networks. Even though clustering uses TDMA (Time Division Multiple Access) for avoiding congestion control for intra clustering data transmission, but it may fail in some critical situation. This paper proposed a energy efficient clustering which avoid data redundancy in a dense sensor network until the network becomes sparse and hence uses the TDMA efficiently during high density of the nodes.

## 1. INTRODUCTION

A sensor network is an infrastructure composed of sensing, computing and communication elements that enable administrators to observe and react to events and phenomena in a specified environment [1]. Wireless Sensor Networks (WSNs) are used in air-traffic control, battlefield management, commercial application, food safety, intrusion detection and many other fields. Each sensor node in a WSN is composed of a sensor, an A/D converter, a processor with memory (e.g., a Strong ARM SA-1100 processor), RF circuits, a DC converter and a battery. Each sensor node has a very limited energy supply and may not be rechargeable. In some cases, nodes may be self-powered. Energy is used for sensing, data processing and communication. Communication is the most energy-intensive process, consuming more energy than the computation process in the operating system. Therefore, a sensor node must be designed intelligently taking all its aspects into consideration, including architecture, algorithms, protocols and circuitry.

Clustering and Multi-Level Clustering methods are used to extend the lifetime of a wireless sensor network. They concentrate more on energy optimization but the main problem is in handling the sensed data. So there may be unnecessary redundant data transmission which may affect the performance of the WSNs

The current study is focused on the design of an algorithm for Clustering and Multi-Level Clustering that can prolong the lifetime of a WSN with avoiding the redundant data transmission and reducing the energy which are wasted for this type of communication.

## 2. RELATED WORK

In this section we are presenting the analysis and study of prevailing clustering and multi level clustering algorithms, in order to analyse the different context of data redundant and energy related issues. Considering these issues we are proposing QBEEC algorithm which reduces the redundant information passing and enhances the quality of information and energy consumed.

### 2.1. Clustering

The first study of cluster-based networks was carried out by (Heinemann et al., 2000). They proposed a Low-energy Adaptive Clustering Hierarchy (LEACH) [3], a routing protocol with self-organizing and adaptive clustering behaviour that randomly distributes the energy load among the sensors in a network. LEACH uses localized coordination to enable scalability and robustness for active networks. It also uses data fusion in the network to reduce the amount of information that must be sent to the sink node.

LEACH used an efficient solution in medium access layer, called time division multiple access (TDMA). TDMA [4] is a well-known medium access control (MAC) protocol, widely utilized in sensor networks. In TDMA, all nodes have a unique timeslot for transmission, which allows collision-free communication and is also very power efficient: since nodes know the scheduled time intervals of communication, they need to be awake only when sending and receiving messages, and can be asleep in the rest of the time; there is no need for idle listening.

LEACH selects the cluster head by using the threshold probability T (n):

$$T(n) = \begin{cases} \dfrac{p}{1 - p(r \bmod(1/p))} & n \in G \\ 0 & \text{others} \end{cases}$$

After the cluster head selection each cluster head will broadcast an advertisement indicating the other non cluster nodes to become its member. Based on the signal strength each non cluster head will join the suitable clusters. Then in the allocated time slot it communicates with its head.
There are other clustering algorithms like FSCH [6], EBHC [7], and SEP [8] which follow the same basis as of LEACH.

In the cases of these algorithms, cluster Head and member communication set up may affect the performance of the network in dense WSNs in the following cases:

*Case 1: Redundant Data Transfer*

There might be a transfer of data to cluster head that is already transferred to the cluster head by its neighbour member and there may be another chance that the same data might have been transmitted by its non member neighbour to another cluster head.

*Case 2: Data Aggregation*

As and when there is an increase in data packets, there is a need of more data aggregation at the Cluster side. So cluster head might have to use its energy to process the redundant data.

*Case 3: Collision*

The emphasis on the selection of the Cluster head is based on probability. If the cluster heads contain many nodes in its cluster then the TDMA cannot guaranty the efficient transmission, because of more members the timeslot will be very less. The node may not get sufficient time to send its data. And some node may waste the allocated time slot to send redundant data. So there is a need of avoiding redundant data that is transmitted and concentrating on QoS of WSNs [5]

## 2.2. Multi Level Clustering

Multi Level Clustering also has a redundant data problem but it is a better alternative instead of single level clustering to reduce power consumption in the sensor networks. There are many Multi Level clustering like HEED [9], PEGASIS [10], and EELBCRP [11] which focus on power optimization. In order to analyse the different context of data redundant in Multi Level Clustering we considered a latest EELBCRP algorithm.

*EELBCRP*: Energy Efficient Level Based Clustering Routing Protocol. This protocol is divided into three phases, setup phase, cluster setup phase and inter cluster routing phase.

*Setup phase*

On the initial deployment, the base station (BS) transmits a level-1 signal with minimum power level. All nodes, which hear this message, set their level as 1. After that, the base station increases its signal power to attain the next level and transmit a level-2 signal. All the nodes that receive the massage but do not set the previous level set their level as 2.

*Cluster setup phase*

In this phase, each level is divided into clusters. The operation of cluster-setup phase is the same as LEACH [3] except the difference of threshold formula. For each level i, each node decide whether or not to become a cluster head for the current round by choosing a random number x between 0 and 1.The node becomes a cluster head for the current round if this number is less than the threshold T(n). The threshold defined as.

$$T_i(n) = \begin{cases} \frac{P \times c}{1 - P \times (r \bmod \frac{1}{P})} \times \left(\frac{U_i - d(n, BS)}{U_i - L_i}\right) \times \left(\frac{E_{cur}(n)}{E_{ini}(n)}\right)^k & \text{if } n \in Z \\ 0 & \text{Otherwise} \end{cases}$$

**Inter cluster routing**

After the cluster formation, the cluster heads broadcast the aggregate data to the next level. At the next level, the nodes aggregate their data and sends to their cluster heads. In this manner the cluster heads at the last level transmit the final information to the BS.

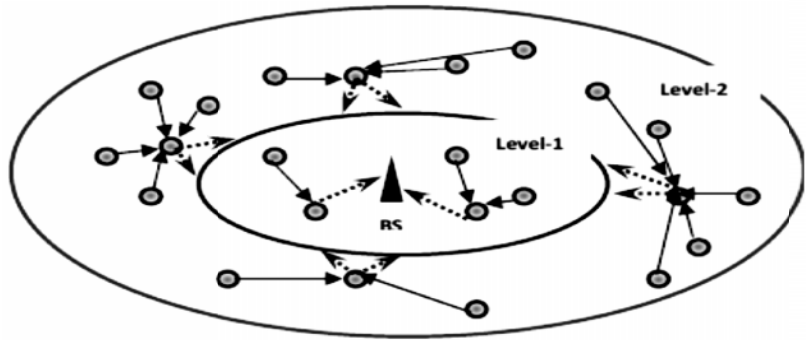

**Fig. 1** EELBCRP Clustering

*Data Redundant Issue in EELBCRP:* In the Inner Clustering Routing the cluster head of lower level broadcast the aggregated data to the other nodes of the upper level. There might have chance that the data broadcasted by the lower level may duplicate at the higher level.

The proposed algorithm aims to improve the QoI (Quality of Information) and Power Optimization by avoiding the redundant data based on the node density.

## 3. QIBEEC ALGORITHM

The proposed QIBEEC (Quality of Information Based Energy Efficient Clustering) algorithm mechanism is as follows:

*QIBEEC Algorithm:*

**Selection of Cluster Head and Cluster Formation**

Step 1: Get the network parameters value: **A** (Area of Network), **L** (Length of a Network)
Step 2: Assign the energy required to Transmission and Aggregation in **nano Joule** and packet size in **bits**
Step 3: Calculate the Minimum distance, Threshold Distance
Step 4: Calculate Threshold Energy $M_t$ using the results of step 3
Step 5: Calculate the Residual Energy of Each node
Step 6: If the Energy $>= M_t$
      Select the node as Cluster Head & Mark node_type ='C' (Cluster Head)
      Else
      Mark the node_type ='N' (Normal Node)
Step 7: Perform Step 6 and 7 for all nodes

**RDA Algorithm**

Step 1: Find the Sensing Range **Sr** of nodes (for simulation we assumed a small range for low cost device)
Step 2: Read the number of nodes deployed in a simulation area
Step 3: Calculate the Threshold sensing Region T(r) by using Step 2 and 3
Step 4: For i=2-100
      While node_type =='N'

```
            Calculate the T(r)i with respect to their current position
            If T(r)i covers the T(r)i+1
                    T(r)i will receive a Active Status Message from T(r)i+1 and
             N (i) will become Idle
            Else
                    If the Status of Message is not Active the n
                     N (i) will become active
            End
        End
    End
Step 5: End
```

*Description:*

### I. Criteria for becoming a Cluster Head

The nodes which decided to become a cluster head must have some minimum amount of energy so that it can transmit the packet successfully to the base station. In a typical condition there is a chance of unsuccessful data transmission, because a node might have some minimum energy that is sufficient to receive the data from receiver but may fail to transmit it to base station. This typical condition may also misuse the minimum amounts of energy present in the node instead it could be served as a member and utilize the energy efficiently.

This algorithm aims to calculate a threshold distance and the threshold energy for the node to become Cluster head so that only the node satisfying this threshold energy and distance will become a cluster head.
Equation below gives the threshold distance and energy:

$$M = (\varepsilon_{tx} \quad \varepsilon_{da}) \quad k + \varepsilon_{fs} \quad k(T_d \quad T_d)$$

Where,

$T_d = \overline{M_d{}^2 + M_d{}^2}$ = Threshold distance

$M_d = \frac{A}{4L}$  Where A= area of a Square network, L = Length of a Network area.

$\varepsilon_{tx}$ = Transmitting energy.
$\varepsilon_{da}$ = Data Aggregation Energy
$k$ = Packet size

### II. Formation of Cluster

If the node satisfies the first condition then it is eligible to become a cluster head.

*Selection of Cluster:* In this phase, each cluster head will form a cluster of nodes within its range. The operation of this cluster-setup phase is the same as LEACH.

### III. Avoiding the Data Redundancy

In this phase we assume that all nodes have same sensing range and the nodes are deployed in a square network area where base station in placed at the centre of the network. When the communication occurs every node will first check whether it is in any sensing range of another node that is already active, if so the node will go to sleep mode and no Time slot is allocated to that node so that the current time slot could be allocated to other nodes in a same cluster.

The members will also check the same condition with its cluster head since cluster head is also a node which as same sensing range then the nodes will receive a message from its cluster head to go to idle instead receiving the duplicate data from its members. The nodes decide whether to become idle or active by the threshold range, T(r):

$$T(r) = \frac{Sr}{N}(L)$$

Where,
**Sr** = Sensing Range of the nodes
**N**=Number of Nodes
**L**=Length of Network area

The node will become idle if it is in T(r) of another node or else active. And this operation will continue till the network become sparse after that all nodes will be active in the cluster. The sparse network is calculated based on the number of nodes dead during the operation. This mechanism is done in random way such that no nodes will be idle for many rounds, as the node which is idle in current round can become an active member or a cluster head in the next round or time of operation.

This way of clustering reduces the traffic in intra clustering communication process and all nodes will get sufficient time slot to transmit the data completely to its cluster head without any redundancy.

*Single Level QIBEEC*

This mechanism is applied for Single and Multi level Clustering; single level clustering uses the same network model as LEACH.
Fig.2 shows the QoI based clustering in that we can observe that the nodes which have dark links are active members of their respective cluster, in that we made the nodes idle which are in the sensing range of other nodes

*Multi Level QIBEEC*

The main motivation for moving to multi level is to reduce the communication distance and communication overhead near the base station.

Fig.3 shows the 2 level Multi Hop Clustering where the Formation of cluster is done as same in single level clustering. The communication between the first level and second level cluster head is based on the shortest distance, if one node become cluster head in higher level then it look for its nearest cluster head in the lower level. In the same figure we can observe the square nodes are second level cluster head and big circled nodes are Cluster heads of first level. The small dark nodes represent the active nodes and remaining nodes are idle.

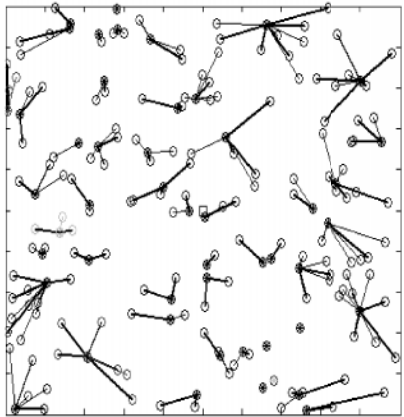
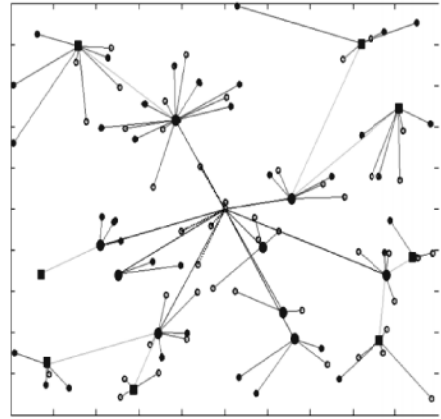

**Fig.: 2**-Single Level QIBEEC     **Fig. 3**- Two Level QIBEEC

## 4. RESULTS

QIBEEC Algorithm mainly aims at increasing the efficiency of WSNs by concentrating on both QoI and energy utilization. The simulation is carried out for analysis for time duration of 200 rounds. The results obtained for the simulated WSNs considering an area of 100X100m with 200 nodes and the sensing range of 15m is taken as per the node capacity, the sensing range will not directly affect the operation as it depends on the density of the nodes deployed also .

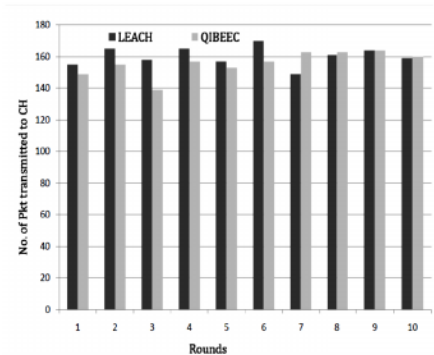
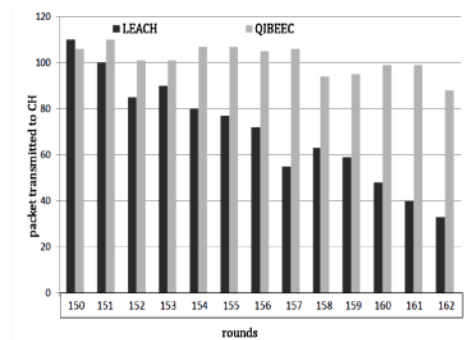

**Fig. 4**-Avg Packet Received at the Cluster      **Fig. 5**- Avg Packet Received at the
Heads during the initial rounds (dense n/w)       Cluster Heads after 150 rounds (sparse case)

For the initial set of 10 rounds of, the network will be dense and the result of the algorithm is shown in Fig. 4. The graph depicts the average number of packet sent by the members to their cluster heads in the initial stages. The packet transmitted in LEACH is more than QoI based clustering. The reason for this might be that LEACH may transmit packets containing similar types of information that is sent by its nearby members, may have duplicate data.

As the QIBEEC in Fig.5 shows the portray result of packet transmission by the members to their cluster heads at the final stage of simulation. The graph signifying that the packet transmitted by the QoI based clustering is more than the LEACH because most of the energy in the LEACH might have been wasted in transmitting redundant data at the initial stage of the transmission. QIBEEC based clustering aims to minimize the unnecessary data transmission at the initial stage and use the same energy at the final stage when the network becomes sparse to avoid the network from rapid increase of failure nodes.

Fig.6 shows the lifetime of a sensor networks where the 1-100 rounds portray the lifetime in dense case and post half life the graph depicts the lifetime of sparse case. In this we can observe that the failure nodes in LEACH at the end are 198 out of 200 where as QIBEEC single level clustering as 178 failure nodes. So QIBEEC based clustering has 12% less than the LEACH.
The result also contains the result of 2 levels clustering which more efficient than both single level clustering hence one can use multi level clustering for better energy efficient along with our proposed QIBEC for good QoI.

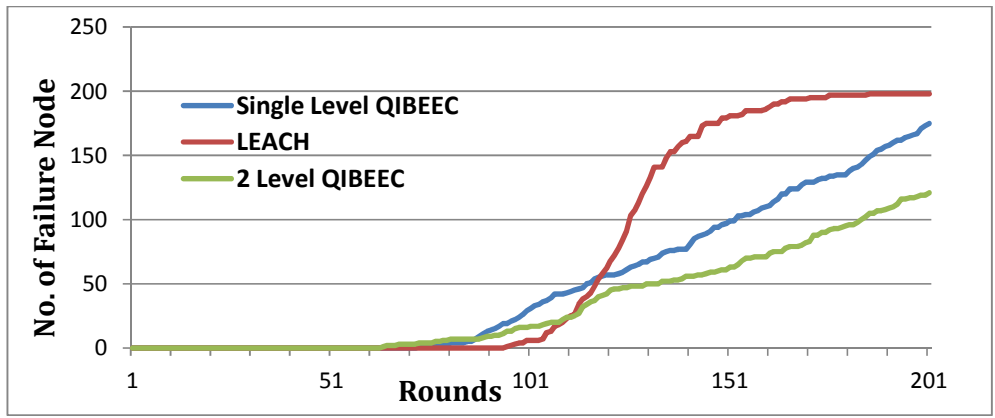

**Figure: 6**- Lifetime of a Network

## 5. CONCLUSIONS

In this paper a new approach for improving a QIBEEC algorithm for wireless sensor networks is proposed. This approaches concentrated on both performance metrics and efficient energy utilization of the wireless sensor networks. The data redundant is one of the main issue in the dense sensor network, so this approach as identified a solution for this issue until the network become sparse. In this we had given solution to both single and multi level clustering for efficient energy utilization and achieving good QoI in the sensor networks

## BIOGRAPH

**ManjuPrasad** received his B.E degree in Computer Science from Visvesvaraya Technological University; Karnataka in 2011.He worked has a Research Assistant in the area of WSNs for 1.5 years and currently pursuing Ph.D (Computer Science) in Visvesvaraya Technological University. His current research interest includes Wireless Sensor Networks.

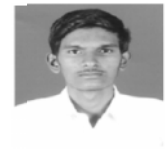

**Andhe Dharani** received her Ph.D from Mother Theresa Women's University, Kodaikanal, India in 2010, currently working as Associate Professor in Department of MCA R.V College of Engineering.  Her current research interest includes Image Processing, Wireless Sensor Networks and undertaken a research project on WSNs. And she is a Lifetime Member of CSI and ISTE.

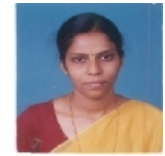